\documentclass[aps, prb, twocolumn,floatfix]{revtex4-2}
\usepackage[colorlinks=true,citecolor=blue,linkcolor=blue,urlcolor=blue]{hyperref}

\usepackage[sort&compress]{natbib}
\usepackage{graphicx,times}
\usepackage{color}

\usepackage{sansmath}
\usepackage{amssymb}

\usepackage{subeqnarray}
\usepackage{diagbox}
\usepackage{makecell}
\usepackage{amsmath}
\usepackage{mathtools}
\usepackage{empheq}
\usepackage{cancel}
\usepackage{bbding}

\newcommand{\msf}[1]{\mathsf{#1}}
\newcommand{\mcl}[1]{\mathcal{#1}}
\newcommand{\bsb}[1]{\boldsymbol{#1}}
\newcommand{\mbf}[1]{\mathbf{#1}}
\newcommand{\mathbbmm}[1]{\text{\usefont{U}{bbm}{m}{n}#1}}

\DeclareSymbolFont{greekletters}{OML}{FiraSans}{m}{n}

\DeclareMathOperator{\msA}{\msf{A}}
\DeclareMathOperator{\msB}{\msf{B}}
\DeclareMathOperator{\msC}{\msf{C}}

\DeclareMathOperator{\msJ}{\msf{J}}
\DeclareMathOperator{\msT}{\msf{T}}
\DeclareMathOperator{\msR}{\msf{R}}
\DeclareMathOperator{\msY}{\msf{Y}}
\DeclareMathOperator{\msZ}{\msf{Z}}
\DeclareMathOperator{\msS}{\msf{S}}
\DeclareMathOperator{\msW}{\msf{W}}

\DeclareMathOperator{\bPhi}{\bsb{\Phi}}
\DeclareMathOperator{\bQ}{\bsb{Q}}
\DeclareMathOperator{\bPi}{\bsb{\Pi}}
\DeclareMathOperator{\bpi}{\bsb{\pi}}

\DeclareMathOperator{\bX}{\bsb{X}}

\DeclareMathOperator{\bxi}{\bsb{\xi}}

\newcommand{\mR}{\mathbbmm{R}}
\newcommand{\mone}{\mathbbmm{1}}

\usepackage{svg}
\newcommand{\orcid}[1]{\href{https://orcid.org/#1}{\includegraphics[width=8pt]{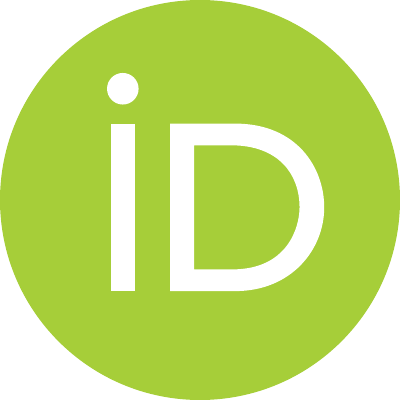}}}
\begin{document}

\title{Quantum fluctuations in electrical multiport linear systems}

\author{A. Parra-Rodriguez\,\orcid{0000-0002-0896-9452}}
\email{ adrian.parra.rodriguez@gmail.com \Envelope}
\affiliation{Department of Physics, University of the Basque Country UPV/EHU, Apartado 644, 48080 Bilbao, Spain}
\affiliation{Institut Quantique and D\'epartement de Physique, Universit\'e de Sherbrooke, Sherbrooke, Qu\'ebec J1K 2R1, Canada.}
\author{I. L. Egusquiza\,\orcid{0000-0002-5827-8027}}
\affiliation{Department of Physics, University of the Basque Country UPV/EHU, Apartado 644, 48080 Bilbao, Spain}

\begin{abstract}
  We present an extension of the classical Johnson-Nyquist theorem for multiport classical electrical passive linear networks by [Twiss, J. Appl. Phys. {\bf 26}, 599 (1955)] to the quantum case. Conversely, we extend the quantum fluctuation-dissipation result for one port electrical systems to the multiport case, both reciprocal and nonreciprocal, performing a detailed quantum analysis of the canonical Foster lossless immitance expansions. Our results are extended to lossy systems by depicting resistive components as continuous limits of purely lossless lumped-element networks. Simple circuit examples are analyzed, including a linear system lacking a direct impedance representation.
\end{abstract}

\maketitle

\section{Introduction}
Electrical circuit models are regularly used to describe a plethora of phenomena both in the classical~\cite{Pozar:2009} and quantum regimes~\cite{Devoret:2013}. Lately, they have become a fundamental tool in the design of quantum processors made of superconducting materials, upon which quantum information protocols are run~\cite{Arute:2019}. Amongst their immense applications, electric lumped-element circuits have been employed to understand  fundamental aspects of noise, be it classical~\cite{Schremp:1949} or quantum in nature~\cite{Feynman:1963}. 

In essence, the superconducting chips that form  the cores of the biggest quantum processors~\cite{Arute:2019,Wu:2021} involve linear and nonlinear systems in mutual interaction. A very common scenario is that of multiple Josephson-junction-based qubits embedded in a linear (possibly nonreciprocal) environment. There, the nonlinear units define input-output ports to complex electromagnetic surroundings. Thus, the growing need for multiport analysis has become a reality. A variety of solutions have been proposed~\cite{Nigg:2012,Solgun:2015,Minev:2020,ParraRodriguez:2019}, and yet further analysis is required.

On a different line, nonreciprocity is expected to become a crucial element for future quantum computers, and is of fundamental interest by itself in the context of superconducting circuits. Some steps towards the implementation of non reciprocal (and thus necessarily multiport) elements in the quantum regime have already been taken~\cite{Sliwa:2015,Kerckhoff:2015,Chapman:2017,Mahoney:2017,Barzanjeh:2017}. These elements are very much desired because they allow non-trivial quantum information directionality in a chip and are part of the solution to the frequency-crowding problem~\cite{Schutjens:2013,Reagor:2018}.

For any kind of device that works in a quantum regime, be it reciprocal or non reciprocal, one-port or multiport,  we require a clear view of how  quantum fluctuations appear and are to be described. At a minimum, a theoretical description of fluctuations in the ideal models used for these devices is  necessary to understand their possible fundamental limitations.

To date, however, the strongest fundamental result on quantum fluctuations in (superconducting) electrical circuits is the one presented by Devoret for the one port case~\cite{Devoret:1997,Vool:2017}. On the classical side, Twiss~\cite{Twiss:1955} gave a non reciprocal extention of the Johnson-Nyquist noise theorem~\cite{Johnson:1928,Nyquist:1928}. Regarding the need for multiport elements in superconducting circuits, Solgun {\it et al.}~\cite{Solgun:2019} have constructed effective Hamiltonian descriptions for qubits coupled to reciprocal multiport environments.

In this article we weave together these three strands of inquiry, i.e. multiport analysis, nonreciprocity, and the study of quantum fluctuations,  by extending the Johnson-Nyquist noise theorem for electrical multiport nonreciprocal linear systems presented by Twiss~\cite{Twiss:1955} to the quantum regime in the spirit of Devoret~\cite{Devoret:1997,Vool:2017} by making use of the multiport version~\cite{Newcomb:1966} of Foster's reactance theorem~\cite{Foster:1924}. Such models are commonly used to describe various types of quantum noise~\cite{Gardiner:2000,Clerk:2010} when a continuous infinite-limit number of harmonic oscillators is taken, ideas first presented and used by Feyman and Vernon~\cite{Feynman:1963} and Caldeira and Leggett~\cite{CaldeiraLeggett:1983}. Even though our presentation uses the language of electrical circuits, its applicability extends to all quantum passive linear systems.

The results here presented rely on the partial fraction decomposition of causal lossless linear responses described by complex matrix functions with a discrete set of poles, see~\cite{Solgun:2014,Solgun:2015,Russer:2012,ParraRodriguez:2018} for related quantum network analysis of lossless and lossy linear systems. A proof based on the exact solution of Heisenberg's equations is derived in Appendix~\ref{App:LinearProof}. Under  standard conditions for the existence of a continuous limit of harmonic resonant frequencies, the result extends to linear systems with energy losses. 

The article is structured as follows. In Sec.~\ref{Sec:GenFor} we present the main results of the article, i.e., the general formulae for computing two-point correlators of flux and charge variables in general linear systems. We verify the recipe in Sec.~\ref{Sec:NRHO} for the fundamental two-port nonreciprocal harmonic oscillator comparing it with a direct computation. The proof is extended to more general lossless systems in Sec.~\ref{Sec:PLLS} by making use of the multiport Foster expansions. In Sec.~\ref{Sec:Lossy}, we argue the use of the formulae in the context of lossy systems, considered as continuous limits of lossless responses with infinite number of poles. In Sec.~\ref{Sec:Singular_Smat}, we exemplify the main results by computing a flux-flux correlator for a singular network for which no direct impedance response is at hand. We finish with conclusions and a perspective on future work in Sec.~\ref{Sec:Conclusions}.

\section{General Formulae for Linear Systems}
\label{Sec:GenFor}
Equilibrium fluctuation-dissipation relations on linear circuits are usually obtained from immitance matrices, yet not all general linear systems accept such a direct description. This manuscript provides a method to obtain generic relations, within the context of electric circuits, that can be easily generalized to other physical linear systems.

In electric circuits, a multiport linear device can be always described by its scattering matrix parameters  $\msf{S}(s)$ \cite{Pozar:2009} where $s\in\mathbbmm{C}$ (written in Laplace space, with $f(s)=\int_{0}^{\infty}f(t)e^{-st}dt$), i.e., a matrix relating voltages ($V$) and currents ($I$) at its ports $\bsb{b}=\msf{S}\bsb{a}$, where $b_k=(V_k-Z_k^* I_k)/\sqrt{\Re\{Z_k\}}$ and $a_k=(V_k+Z_k I_k)/)/\sqrt{\Re\{Z_k\}}$ are output and input signals at port $k$, respectively. Without loss of generality, we take the reference impedances to be homogeneous and real, $Z_k=R\in \mathbb{R}$. We denote the number of ports by $N$ in what follows.

The scattering response encodes all the information about the system, e.g. a network is lossless when $\msf{S}$ is unitary, or reciprocal (time-reversal invariant) when $\msf{S}=\msf{S}^T$. Other common descriptions of multiport linear systems are the impedance $\msZ=R(1-\msf{S})^{-1}(1+\msf{S})$ and admittance  $\msf{Y}=\msf{Z}^{-1}$ matrices that relate voltages and currents, or fluxes and charges at the output ports as $\bPhi(s)=\msf{Z}(s)\bQ(s)$~\cite{Pozar:2009}. Here, and in the rest of the article, we assume charges and fluxes at initial times to be zero. It is however well known that immittance descriptions of linear devices, reciprocal or not, do not always exist, so that working with $\msf{S}$ is sometimes unavoidable \cite{Carlin:1964,Pozar:2009}. Mathematically, this is a consequence of voltage and/or current constraints at the ports, and it happens whenever the $\mathsf{S}$ matrix has $+1$ and $-1$ eigenvalues, corresponding to very particular phase differences in signals passing through the system from port $i$ to port $j$. For example, ideal circulators with even number of ports, even number of ``$+1$" and ``$-1$" entries admit only an $\msS$ representation~\cite{ParraRodriguez:2019}. These situations can be handled properly and exactly, as described in section \ref{Sec:Singular_Smat}, due to the universal equivalence of, on the hand, scattering responses with, on the other,  lower-rank immitances filtered by a network of ideal transformers. In what follows we assume that such an  analysis has been already performed.
 
\begin{figure}[h!]
	\centering\includegraphics[width=1\linewidth]{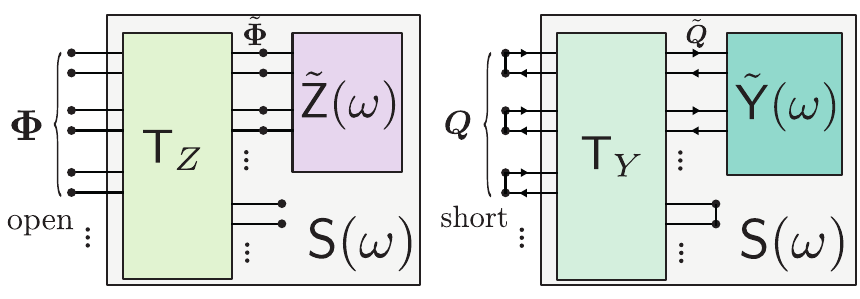}
	\caption{\label{fig:Qfluctuations_S} Representations of linear systems in terms of scattering matrices can always be described in terms of an initial network of ideal (Belevitch) transformers ($\msT$) and an (a) impedance $\msZ$ or (b) admittance $\msY$ response.}
\end{figure}

The main result of this article is the proof that flux and charge fluctuations at the ports of causal linear (nonreciprocal) quantum systems at thermal equilibrium are determined by the immitance matrix responses in the form 
\begin{align}
&\langle\bPhi(t)\bPhi(0)^T\rangle_{\text{th}}=\frac{\hbar}{\pi}\int_{\mR}\frac{d\omega}{\omega}\left[ n_{\mathrm{th}}(\omega)+1\right]\bar{\msZ}^H(\omega)e^{-i\omega t},\label{eq:Phit_Phi0_GF}\\
&\langle \bQ(t)\bQ(0)^T\rangle_{\text{th}}=\frac{\hbar}{\pi}\int_{\mR}\frac{d\omega}{\omega}\left[ n_{\mathrm{th}}(\omega)+1\right]\bar{\msY}^H(\omega)e^{-i\omega t},\label{eq:Qt_Q0_GF}
\end{align}
for open and shorted ports, respectively. Here $\mbf{a}^T=(a_1,\dots)$,  $\bar{\msf{F}}=\msf{P}_F^T\tilde{\msf{F}}\msf{P}_F$  and $\tilde{\msf{F}}\in\{\tilde{\msZ},\tilde{\msY}\}$ are the filtered and lower-rank immitance response matrices, respectively. $\msf{P}_F$ are (possibly) rectangular submatrices of the  full-rank square transformer matrices $\msT_{F}$, see Fig.~\ref{fig:Qfluctuations_S}. 

The full-rank immitance matrices are connected to output ports through ideal Belevitch transformers~\cite{Belevitch:1950}. Thus, for the reduced impedance description we have the relation   $\bPhi=\msT_{Z}^T\tilde{\bPhi}$, where $\bPhi$ and $\tilde{\bPhi}$ are the external and internal flux vectors, respectively, of non-trivial (non-zero) dimensions $N$ and $N-k$ ($0\leq k < N$). Analogously, the internal and external loop charges are related as  $\bQ=-\msT_{Y}^T\tilde{\bQ}$ for the reduced admittance description. 
$\msf{P}_F$ are constructed from the $N-k$ rows of their corresponding orthogonal transformation matrix~\cite{Newcomb:1966}. In what follows, we do not use the tilde/bar symbols (nor are the transformers required) for immittance matrices of same rank as their scattering responses.

Linear systems representable by their immitance responses do not require such ideal transformer networks, i.e.,  $\msf{P}_F, \msT_F\rightarrow\mone_N$, see Ref.~\cite{Newcomb:1966} for further details. The hermitian part of the immitance matrices in the general formulae (\ref{eq:Phit_Phi0_GF}) and (\ref{eq:Qt_Q0_GF}) is defined as $\msf{F}_{ij}^H=(\msf{F}_{ij}+\msf{F}_{ji}^*)/2$, and we have also used the standard number thermal distribution $n_{\mathrm{th}}(\omega)=\left(\coth\left(\beta\hbar\omega/2\right)-1\right)/2$.

We recall that the immittance responses appearing there are the causal Fourier representations of the linear systems ($s=-i\omega+ 0^+$), i.e., $\msZ(\omega)\equiv\lim\limits_{s\rightarrow-i\omega + 0^+} \msZ(s)$, where the abuse of notation must be properly understood. We further remind the reader that simple poles at the origin and infinite frequency do not contribute to the above expressions given that the open/short conditions at the ports implies that only nontrivial loops, i.e., those formed by two components of the triad of capacitors, inductors and/or gyrators, are accounted for, see below. 

Finally, we remark that the conjugated $\langle \bPi(t)\bPi(0)^T\rangle$ ($\langle \bpi(t) \bpi(0)^T\rangle$) and cross $\langle \bPhi(t) \bPi(0)^T\rangle$ ($\langle \bQ(t) \bpi(0)^T\rangle$) correlations to the accessible open node-flux (short loop-charge) variables can be also easily computed in terms of more general immitance-gain matrices~\cite{Nazarov:2009}, where $\bPi\leftrightarrow\bPhi$ ($\bpi\leftrightarrow\bQ$) are the conjugated charges (fluxes) to the node-flux (loop-charge) variables, see Appendix. It must be remarked, however, that such conjugated variables are not trivially accessible from the output port electrical variables.

\section{Nonreciprocal Harmonic Oscillator}
\label{Sec:NRHO}
\begin{figure}[h]
	\centering\includegraphics[width=1\linewidth]{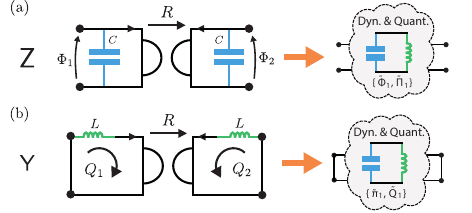}
	\caption{\label{fig:NR_HOs} Fundamental nonreciprocal circuits for the (a) impedance and (b) admittance representations described by configuration-space coordinates $\Phi_i$ and $Q_i$, and pairs of quantized conjugated phase-space variables $\tilde{\Phi}_1\leftrightarrow\tilde{\Pi}_1$ and $\tilde{Q}_1\leftrightarrow\tilde{\pi}_1$, respectively.}
\end{figure}

The core of the proof resides in the analysis of the canonical quantization procedure for the two-port nonreciprocal harmonic oscillator (NR HO) circuit, which plays an analogous role to the LC-oscillator for one-port devices~\cite{Devoret:1997,Vool:2017}. As a consequence of the multiport Foster expansion theorem~\cite{Newcomb:1966}, a general lossless multiport linear response can always be decomposed into weigthed contributions of reciprocal and nonreciprocal harmonic oscillators, as it will be seen in the following section. Thus, let us first focus on the fundamental circuits implementing only simple pairs of poles of a nonreciprocal two-port system in their impedance (admittance) response, which can be represented by a parallel (series) connection of capacitors (inductors) and a gyrator~\cite{Tellegen:1948}, see Figs.~\ref{fig:NR_HOs}(a) and \ref{fig:NR_HOs}(b). We remind the reader that the gyrator represents the fundamental two-port ideal nonreciprocal lumped element implementing lossless time-reversal symmetry breaking in the form of constraints between voltages ($\dot{\bPhi}$) and currents ($\dot{\bQ}$) at its ports, i.e. $\dot{\bQ}=\msY\dot{\bPhi}$, where the admittance matrix is
\begin{align}
	\msY=\frac{1}{R}\begin{pmatrix}
		0&1\\-1&0
	\end{pmatrix}.
\end{align}
Realistic gyrators and other more complex nonreciprocal devices like circulators are being very intensely researched, with theoretical and experimental efforts based on active (driven nonlinearities)~\cite{Barzanjeh:2017,Kerckhoff:2015,Sliwa:2015,Chapman:2017}, and passive circuits (Hall effect, magnetically-biased Josephson rings)~\cite{Koch:2010,Viola:2014,Mahoney:2017,Mueller:2018}. Due to the current-voltage mixing constraints of the gyrator, a parallel capacitor in one port is {\it seen} as a series inductor on the other, and vice-versa~\cite{Tellegen:1948,Newcomb:1966,Anderson:1975}. Thus, the circuits in Fig.~\ref{fig:NR_HOs} implement nonreciprocal harmonic oscillators~\cite{ParraRodriguezPhD:2021}. These circuital representations will allow us to compute open- (short-) port flux (charge) fluctuations. It must be emphasized that fluctuations of conjugated variables to the node fluxes $\Phi_i$ (resp. loop charges $Q_i$) in Fig~\ref{fig:NR_HOs} throught the use of Heisenberg equations do not match those computed with the dual formulae (\ref{eq:Qt_Q0_GF}) (resp. (\ref{eq:Phit_Phi0_GF}), due to the open- (short-) circuit condition not being respected. 

Making use of the multiport Foster immitance expansion theorems~\cite{Newcomb:1966}, we will generalize the results of this section to general linear systems down below. Going beyond discrete systems, careful continuous limits may be taken to derive meaningful results for lossy networks that connect with Johnson-Nyquist-Twiss classical noise formulae~\cite{Twiss:1955}. 

Let us begin our study by computing flux fluctuations for the NR HO circuit depicted in Fig.~\ref{fig:NR_HOs}(a). The Laplace transform of its impedance response reads 
\begin{align}
\msf{Z}_\Omega(s)&=\frac{1}{s^2+\Omega^2}\begin{pmatrix}
s/C&-R\Omega^2\\R\Omega^2&s/ C
\end{pmatrix},\label{eq:Zs_NR-HO}
\end{align} 
with $\Omega=1/R C$, and a possible Lagrangian description written in terms of the flux variables at the active nodes is~\cite{ParraRodriguezPhD:2021}
\begin{equation}
	L=\frac{C}{2}\left(\dot{\Phi}_1^2+\dot{\Phi}_2^2\right)-\frac{1}{2R}\left[\dot{\Phi}_2\Phi_1-\dot{\Phi}_1\Phi_2\right],\label{eq:Lagrangian_CCG}
\end{equation}
where $R$ is the fundamental resistance gyration parameter describing the gyrator, the minimal device breaking time reversal symmetry~\cite{Tellegen:1948}. Observe the formal similarity with another system in which time reversal invariance is broken, namely the Landau problem~\cite{Egusquiza:2022,Rymarz:2021}. A dual analysis can be done for Fig.~\ref{fig:NR_HOs}(b) with  loop-charge variables~\cite{Ulrich:2016}. Observe that the constitutive equations of a gyrator are in fact a constraint. However, this constraint cannot be properly expressed in the configuration space of the two flux variables, nor in the dual configuration space (with charge variables). As we have shown previously~\cite{ParraRodriguez:2019,ParraRodriguezPhD:2021,ParraRodriguez:2022}, this obstacle in the description of nonreciprocal systems can be overcome by using the redundant configuration space above, and eliminating nondynamical variables from the Hamiltonian description. Namely, the full-rank kinetic (capacitive) matrix in Lagrangian \eqref{eq:Lagrangian_CCG} allows for  a Legendre transformation $\Pi_i=\partial L/\partial \dot{\Phi}_i$  to obtain the Hamiltonian 
\begin{align}
	\tilde{H}=\frac{(\Pi_1-\Phi_2/2R)^2}{2C}+\frac{\left(\Pi_2+\Phi_1/2R\right)^2}{2C}.
\end{align}
We can now apply a symplectic transformation $\mathsf{U}$ of the two pairs of conjugate variables, $\tilde{\bX}=\msf{U}\bX$, where $\bX=(\Phi_1,\Phi_2,\Pi_1,\Pi_2)$, 
to reveal its one-oscillator nature 
\begin{equation}
	H=\frac{\tilde{\Pi}_1^2}{2C}+\frac{\tilde{\Phi}_1^2}{2C R^2}\equiv_{\text{quant.}}\hbar \Omega a^\dag a. \label{eq:H_CCG_diag}
      \end{equation}
Again, this is formally identical with the usual analysis of the Landau problem~\cite{Egusquiza:2022,Rymarz:2021}. From this point henceforward, the nondynamical variables $\tilde{\Phi}_2$ and $\tilde{\Pi}_2$ are consistently set to zero, and discarded from the subsequent analysis. Any constant value different from zero would also be possible, and it would amount to a constant shift in the external variables. The Hamiltonian we have obtained, Eq. \eqref{eq:H_CCG_diag}, is susceptible of canonical quantization in the standard manner.

One can compute correlators of the external measurable coordinates,  at the ports, and relate them with the internal degrees of freedom of the system
\begin{align}
	&\langle\Phi_1(t)\Phi_1(0)\rangle_{\text{th}}=\langle\tilde{\Phi}_1(t)\tilde{\Phi}_1(0)\rangle_{\text{th}}=\nonumber\\
	&=\frac{\hbar R}{2}\left[\coth\left(\frac{\beta\hbar\Omega}{2}\right)\cos(\Omega t)-i\sin(\Omega t)\right],\label{eq:corr_phi1_phi1_NRH}
\end{align} which is equal to $\langle\Phi_2(t)\Phi_2(0)\rangle_{\text{th}}$ for symmetry reasons. Equivalently, the cross correlators  $\langle\Phi_1(t)\Phi_2(0)\rangle_{\text{th}}=-R\langle\tilde{\Phi}_1(t)\tilde{\Pi}_1(0)\rangle_{\text{th}}$ of the output port variables linearly depend on position-momentum correlators of the dynamical internal variables $+\pi/2$ out of phase with respect to Eq. (\ref{eq:corr_phi1_phi1_NRH}). 
The identification of Eq. \eqref{eq:corr_phi1_phi1_NRH} amounts, as stated above, to a particular choice of origin of coordinates for the external variables, and is operationally well defined.

It is easy to check now that the above correlators can be directly computed with the general formula (\ref{eq:Phit_Phi0_GF}) through the hermitian part of the causal response associated with (\ref{eq:Zs_NR-HO}), i.e., 
\begin{align}
	\msf{Z}_\Omega^H(\omega)&=\frac{\Omega\pi R}{2}\left[(\mone_2(\delta_{\Omega}+\delta_{-\Omega})+\sigma_y(\delta_{\Omega}-\delta_{-\Omega})\right],\label{eq:Z_H_2P_NRHO}
\end{align}
where $\delta_{\pm\Omega}=\delta(\omega\mp\Omega)$. Correlators for the loop charge variables for the dual circuit in Fig.~\ref{fig:NR_HOs}b can be analogously computed in terms of its admittance response. 

\section{Passive Lossless Linear Systems}
\label{Sec:PLLS}
\begin{figure}[h]
	\centering\includegraphics[width=1\linewidth]{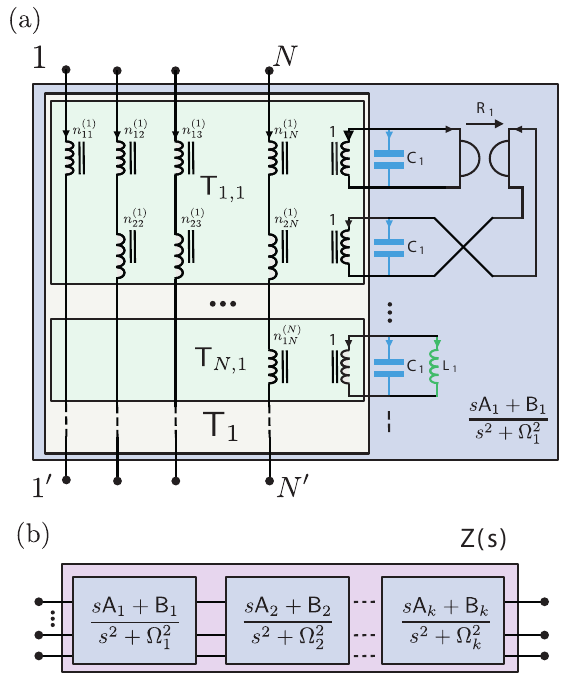}
	\caption{\label{fig:Z_matrix} (a) Canonical realization of a nonreciprocal stage $\msZ_1(s)=\frac{s\mathsf{A}_1+ \mathsf{B}_1}{s^2+\Omega_1^2}$ for (b) a multipole response $\msZ(s)=\sum_k \msZ_k$.} Capacitors and inductors implement $\msA_k$ while nonreciprocal devices realize $\msB_k$. Transformer networks $\msT_k$ generalize the response seen from the output ports. Impedance responses with odd number of ports ($N$) must have at least one reciprocal stage per harmonic frequency.
\end{figure}

Having understood the two-port nonreciprocal harmonic oscillator, it is now easy to generalize the result to multiport general linear responses by using the results in matrix fraction expansions from last century~\cite{Foster:1924,Tellegen:1948, Belevitch:1950,Newcomb:1966}. Analogously to the one port case, any lossless response can be decomposed in weighted contributions from reciprocal and nonreciprocal harmonic oscillators through ideal transformers, see a generic lumped-element circuital realization of a a generic (non-)reciprocal stage in Fig. {\ref{fig:Z_matrix}}.  Such electrical lossless multiport systems can be described in terms of four kinds of lumped circuit elements: capacitors, inductors, nonreciprocal elements (gyrators/circulators) and ideal transformers. The first two elements of the set represent minimal containers of electrical and magnetic energy, respectively,  while the latter ones induce direct or mixing constraints between {\it dual} pairs of quantities such as fluxes (voltages) and charges (currents). While transformer constraints can be systematically eliminated if our configuration space contains just either flux or charge variables, this is not  the case for nonreciprocal constraints. They can, however, always be handled in the form portrayed above, i.e., by redundant configuration spaces and elimination of the nondynamical variables in phase space.

Let us now see how the flux fluctuation formula unfolds for a general passive lossless system. By virtue of the multiport Foster theorem~\cite{Newcomb:1966}, linear systems described by impedance matrices satisfying a causal response (viz. lossless positive-real matrices) can be decomposed as
\begin{align}
		\msf{Z}(s)=\msf{B}_{\infty}+s^{-1}\msf{A}_{0}+s\msf{A}_{\infty}+\sum_{k=1}^{\infty}\frac{s\msf{A}_{k}+\msf{B}_k}{s^{2}+\Omega_{k}^{2}}.\label{eq:NR_MportL_Zmat}
\end{align}
where $s\in\mathbbmm{C}$, and $\msB$ ($\msA$) are (skew-)symmetric matrices, and $\Omega_k$ are a possibly (discrete) infinite set of harmonic resonance frequencies. 

In parallel with the formula for the fluctuations of a one-port linear device~\cite{Devoret:1997,Vool:2017}, only the contributions from the nontrivial loops (those realized with finite-frequency poles, i.e., $\msA_k$ and $\msB_k$) are considered in the general formulae (\ref{eq:Phit_Phi0_GF}) and (\ref{eq:Qt_Q0_GF}). Imposing open boundary conditions, i.e., no external currents flowing in, means that the currents flowing through the first three terms of the right-hand side (RHS) of (\ref{eq:NR_MportL_Zmat}) are zero, and thus they will not contribute to the formula (\ref{eq:Phit_Phi0_GF}). It is worth remarking that this consideration is superfluous for the terms $\msB_\infty$ and $s\msA_\infty$, which are directly discarded on taking the hermitian part of the matrix $\msZ^H(\omega)$, but it is mandatory for the poles at zero frequency ($s^{-1}\msA_0$).

Each reciprocal and/or nonreciprocal stage (with/without $\msB_k$) can be generally fraction-expanded in a sum of simple two-port nonreciprocal and reciprocal stages, connected to Belevitch transformers~\cite{Newcomb:1966}
\begin{align}\label{eq:twoportstage}
	\frac{s\msf{A}_k+\msf{B}_k}{s^{2}+\Omega_k^{2}}=\sum_i \msT_{i,k}^T \msZ_{\Omega_k}(s) \msT_{i,k}
\end{align}
 where the two-by-two impedance matrix $\msZ_{\Omega_k}(s)$ is either a  diagonal one-by-one matrix for a reciprocal harmonic oscillator contribution $Z_{\Omega_k}(s)=(s/C)/(s^2+\Omega_k^2)$, or that of (\ref{eq:Zs_NR-HO}) for a nonreciprocal contribution, where $R_k=1/\Omega_k C_k$. $\msT_{i,k}$ are (possibly rectangular) transformer matrices with $i\leq N$, and $N$ the maximum rank of $\msA_k$ and $\msB_k$. Therefore, we can associate an internal Hamiltonian to each of these two-by-two matrices, as in the case above with one single dynamical variable, see Eq. (\ref{eq:H_CCG_diag}). We aggregate all internal flux fluctuations of the multiport system by the linear relation of flux coordinates on both sides of the ideal transformer matrix and compute correlators for flux variables on the right (inner flux variables $\bPhi_{\mathrm{inner}}$) in terms of those on the left (output port variables $\bPhi$) 
 \begin{align}
 	\langle \bPhi(t)\bPhi^T(0)\rangle=&\msT^T\langle \bPhi_\text{inner}(t)\bPhi^T_\text{inner}(0)\rangle\msT,
 \end{align}
where the matrix $\msT$ is composed from the rows of transformers $\msT_{i,k}$ in all different resonant frequencies. For example, for the one pole impedance matrix in Fig. \ref{fig:Z_matrix}(a), the combined Belevitch transformers are stacked together as 
\begin{align}
	\msT_{1}^T=\left[
							\msT_{1,1}^T\,\,\msT_{2,1}^T\,\,\dots\,\,\msT_{N,1}^T\right],
\end{align}
with $\msT_{i,1}$ having the elements $n^{(i)}_{mn}$ in row $m$ and column $n$. We recover by linearity Eq. (\ref{eq:Phit_Phi0_GF}) decomposing the sum over all frequencies and stages when the impedance matrix directly exists. Given the open boundary conditions at the external ports we just need to add the the weighted independant contributions of all oscillators.
An additional (orthogonal) transformer is required to make the correspondance between inner and outer variables for more general linear systems, see an example below.

A dual analysis can be carried out to derive the general formula for charge fluctuations (\ref{eq:Qt_Q0_GF}) by synthesizing a dual circuit for the admittance response 
\begin{align}
\msf{Y}(s)=\msf{E}_{\infty}+s^{-1}\msf{D}_{0}+s\msf{D}_{\infty}+\sum_{k=1}^{\infty}\frac{s\msf{D}_{k}+\msf{E}_k}{s^{2}+\Omega_{k}^{2}},\label{eq:NR_MportL_Ymat}
\end{align}
i.e., connecting in parallel Belevitch transformers whose secondary ports are directly attached in series to inductors and gyrators, see Chapt. 7 in~\cite{Newcomb:1966}. In an equivalent manner, the first three terms on the right-hand side of (\ref{eq:NR_MportL_Ymat}) will not contribute under the short-circuit condition at the ports.

\section{Lossy Systems}
\label{Sec:Lossy}
The above formulae have been obtained to describe passive lossless linear circuits, but they can be naturally extended to cases where there is presence of energy loss/decoherence effects, by following the standard routine of representing a dissipative causal passive function (with a positive smooth real part) with a continuous limit of an infinite (not unique) sequence of lumped degrees of freedom~\cite{Feynman:1963,CaldeiraLeggett:1983,YurkeDenker:1984} whose causal response functions' real part is a sequence of delta distributions. 

\begin{figure}[h]
	\centering
	\includegraphics[width=.75\linewidth]{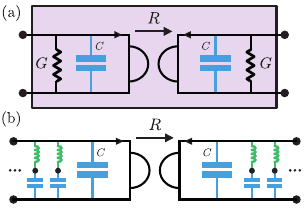}
	\caption{\label{fig:Dissipative_NR_HOs}(a) Ideal dissipative nonreciprocal harmonic oscillator with parallel conductances $G$ connected to the output ports with (b) a  representation of the ideal resistive elements in terms of an infinite continuous series of harmonic oscillators.}
\end{figure}

We can then extend Devoret's argument~\cite{Devoret:1997,Vool:2017} to multiport linear systems to compute, for example, the flux fluctuations at the ports of a dissipative nonreciprocal two-port harmonic oscillator with conductances in parallel, as in Fig.~\ref{fig:Dissipative_NR_HOs}(a), inserting in Eq.  (\ref{eq:Phit_Phi0_GF}) the hermitian part of the impedance matrix of the system $\msZ^H(\omega)=\msZ_+ + \msZ_-$, where 
\begin{align}
	\msZ(s)&=\frac{1/C^2}{(s+ G/C)^2+\Omega^2}\begin{pmatrix}
		G+sC&-1/R\\1/R&G+sC
	\end{pmatrix},\nonumber\\
	\msZ_\pm&=\frac{G(\Omega R)^2/2}{(\omega\pm\Omega)^2+\left(G \Omega R\right)^2}\left(\mone_2\mp\sigma_y\right),\nonumber
\end{align}
and $C=1/\Omega R$. Charge fluctuations conjugate to the flux variables at the ports can be easily computed in terms of the previous flux fluctuations by considering the conductances $G$ to be the continuous limit of a simple sequence of parallel LC-harmonic oscillators~\cite{Devoret:1997,Vool:2017}, see Fig.~\ref{fig:Dissipative_NR_HOs}(b). A Lagrangian representation of that circuit is
\begin{align}
	L=&\,\frac{1}{2} \left(\dot{\bPhi}^T\msC \dot{\bPhi}+\dot{\bPhi}^T\msY\bPhi\right)\nonumber\\
	&+\sum_{\alpha,n}\left[\frac{C_{n}\dot{\phi}_{\alpha, n}^2}{2}-\frac{(\phi_{\alpha, n}-\Phi_\alpha)^2}{2 L_{n}}\right]
\end{align}
with $\phi_{\alpha,n}$ the internal node-flux coordinates of the LC harmonic oscillators for port $\alpha$, and $\msC=\text{diag}(C,C)$. Possible sequences used to approximate the conductance $G$ are $C_{n}=2G/\pi n^2 \Delta \Omega$ and $L_{n}=\pi/2G\Delta \Omega$, where the frequency step $\Delta \Omega\rightarrow0$~\cite{Devoret:1997,Vool:2017}. For this particular circuit, the conjugate charges to the external flux nodes are $\bPi=\partial L/\partial \dot{\bPhi}=\msC\dot{\bPhi}+\msY\bPhi$, such that the cross-port correlator can be computed in terms of the flux correlators, i.e.,  $\langle \Pi_1(t)\Pi_2(0)\rangle = -C^2 \partial_t^2 \langle \Phi_1(t)\Phi_2(0)\rangle-\frac{1}{4R^2}\langle\Phi_1(t)\Phi_2(0)\rangle-\frac{C}{2R}\partial_t \left[\langle\Phi_2(t)\Phi_1(0)\rangle-\langle\Phi_1(t)\Phi_2(0)\rangle\right].$

Let us stress again that these charge fluctuations are not the same as those computed with equation (\ref{eq:Qt_Q0_GF}), where short-circuit boundary conditions at the ports would be assumed (with zero net correlations, i.e., $\langle Q_i(t)Q_j(0)\rangle=0$), hence the use of a different symbol. 

\section{``Singular" scattering matrix}
\label{Sec:Singular_Smat}
Finally, let us use the general formula for computing flux quantum fluctuations in an example with a harmonic LC oscillator embedded in a 3-port scattering matrix without a straight immitance description, i.e., when $+1$ is an eigenvalue of the scattering matrix $\msS=\mone-\frac{2R(s^2+\Omega^2)}{s/C+R(s^2+\Omega^2)}\mbf{d}\mbf{d}^T$, with $\mbf{d}^T=\left(\mcl{C}_z, -\mcl{S}_z\mcl{C}_x,\mcl{S}_z\mcl{S}_x\right)$ and $\mcl{C}_\alpha=\cos(\varphi_\alpha)$ and $\mcl{S}_\alpha=\sin(\varphi_\alpha)$. Given that the system is passive and causal, there always exists an orthogonal transformation $\msT_Z$~\cite{Newcomb:1966} such that
\begin{align}
	\msS&=\msT_Z^T(\mone_2^3+\tilde{\msS})\msT_Z,\label{eq:Smat_singular}
\end{align}
with $\mone_k^n$ the identity matrix of rank $k$ embedded in $n$ dimensions, which play the role of $k$ open ports on the RHS of the orthogonal transformer $\msT_{Z}$. 
\begin{figure}[h]
	\centering
	\includegraphics[width=.75\linewidth]{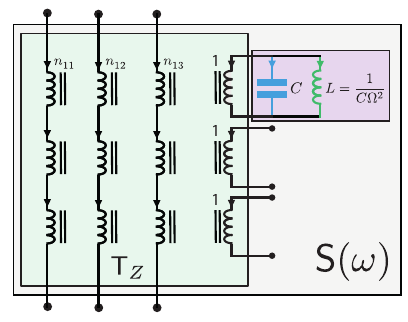}
	\caption{\label{fig:S_matrix_HO}3-port scattering matrix of Eq.  (\ref{eq:Smat_singular}) with two ``$+1$" eigenvalues decomposed into a one-port impedance response ($\tilde{\msZ}$), and a Belevitch transformer ($\msT_Z$).}
\end{figure}

In this example, and for the sake of simplicity, one can choose a basis for the orthogonal space expanded by $\mbf{d}$ such that the transformer is the sequence of rotations in the $z$ and $x$ axes in $\mathbb{R}^3$, i.e., $\msT_Z=\msR_z(\varphi_z)\msR_x(\varphi_x)$. $\tilde{\msS}$ is a scattering matrix of rank 1. Assuming a characteristic reference impedance $R$ at all ports, a reduced ($1\times1$) impedance matrix of a HO is obtained
\begin{align}
	\tilde{\msZ}&=R(1-\tilde{\msS})^{-1}(1+\tilde{\msS})=\frac{s/C}{s^2+\Omega^2}\label{eq:Z_2x2_example}.
\end{align}
The flux fluctuations at the ports are computed directly, inserting the impedance response \emph{seen} by the outer ports
\begin{align}
\bar{\msZ}^H(\omega)=\msf{P}_Z^T\tilde{\msZ}^H(\omega)\msf{P}_Z=\frac{\Omega R\pi}{2}\mbf{d}\mbf{d}^T\left(\delta_{\Omega}+\delta_{-\Omega}\right)
\end{align}
in Eq. (\ref{eq:Phit_Phi0_GF}), where the projector $\msf{P}_Z=\mbf{d}^T=(n_{11},n_{12},n_{13})$  corresponds to the first row of the transformer matrix $\msT_Z$, due to the relation between internal and external fluxes. We recall again that the two open ports on the RHS of the transformer, see Fig.~\ref{fig:Qfluctuations_S}(a), do not contribute to the fluctuations. In this particular example, the charge fluctuations given short-circuit boundary conditions can be immediately computed in terms of its associated admittance matrix because ``$-1$" is not an eigenvalue of $\msS$.

\section{Conclusions and outlook}
\label{Sec:Conclusions}
In this article, we have proved and generalized the computation of quantum fluctuations of conjugated flux and charge variables in multiport linear electrical systems by making use of the multiport Foster expansion of passive causal lossless matrices. The core of the proof resides in the identification of the dynamical (quantized) and nondynamical (discarded) variables for the two-port nonreciprocal harmonic oscillator. 

Our results include the case where the linear system breaks time-reversal symmetry (reciprocity), thus generalizing the classical formulae of Johnson-Nyquist-Twiss, and are applicable to all quantum linear passive systems. The correlators can be further used in the context of lossy nonreciprocal networks under the mapping of the lossy 2-terminal components, i.e., elements with smooth real response functions, with a continuous limit of infinite sequences of harmonic degrees of freedom. Applications of our theorem include but are not restricted to the computation of coherence and decay rates of multi-qubit nonreciprocal superconducting quantum chips~\cite{Devoret:1997,Nigg:2012,Solgun:2014,Solgun:2015}. However, further work will be necessary to bring together the results here presented for multiport linear passive systems, and the fluctuation-dissipation theorems for non-linear systems, such as tunnel-junctions, see~\cite{Parlavecchio:2015,Roussel:2016}.

\begin{acknowledgments}
	I. L. E. acknowledges support from the Basque Government through grant IT986-16.
\end{acknowledgments}

\appendix
\section{Derivation based on Heisenberg's equations}
\label{App:LinearProof}

For the sake of completeness, we present in this appendix the proof of the general formulae for the computation of flux and charge quantum fluctuations for linear passive systems. The first elements of the proof are well known from the reciprocal context; yet we strive to give all the details also of those first steps.

The starting point is that a linear system  and  its immitance responses can be described in terms of quadratic Hamiltonians characterized by Hamiltonian matrices $\mathsf{h}$, as we see below. Next we observe that, algebraically, linearity and passivity imply the restriction to definite semi-positive  Hamiltonian matrices $\msf{h}$. In a well referenced piece of work \cite{Williamson:1936},  Williamson established all the possible canonical forms of Hamiltonian matrices according to symplectic invariance. There he established that these canonical forms correspond to the Jordan canonical forms  of $\mathsf{Jh}$, where $\mathsf{J}$ is the canonical symplectic matrix, see more details on normal forms for positive semidefinite matrices in~\cite{Egusquiza:2022}. For these purposes it is best to organize the variables in such a way that the canonical symplectic matrix $\mathsf{J}$ is in block diagonal form with elementary two-by-two blocks
\begin{equation}
	\label{eq:jmat}
	\mathsf{J}_2=
	\begin{pmatrix}
		0&1\\-1&0
	\end{pmatrix}\,.
\end{equation}

As we have pointed out, we have to consider the case of positive semidefinite Hamiltonian matrices. The existence of zero eigenvalues will be a signal of one or both of two types of variables, nondynamical or free-particle. The original text of Williamson is somewhat obscure in this regard, albeit complete, and a much clearer presentation of this fact will be found in the work of H\"ormander \cite{Hoermander:1995}. The free-particle sector will give rise to poles at the origin (or at infinity) in the corresponding gain-immitance matrices (see below), while the nondynamical sector, if required, will give a constant term.

We are interested in circuital (system) responses under only  open- and short-circuit type conditions, in which case there is no contribution from poles at the origin and at infinity. As a consequence, we consider only Hamiltonian representations without free particle dynamics, i.e., there are no independent capacitor-charge or inductor-flux degrees of freedom. From the point of view of the Hamiltonian matrix $\mathsf{h}$, 
the absence of free particles entails that the canonical form of $\msJ\msf{h}$ does not contain non-diagonal Jordan blocks~\cite{Williamson:1936,Hoermander:1995,Kustura:2019}. 

Coming now to nondynamical degrees of freedom, in the context at hand (electrical circuits) they can only appear from a Lagrangian analysis of the system because of the constraints introduced by nonreciprocal ideal elements. Thus, they will be fixed at constant values; zero, as a matter of fact, by suitable choice of the origin of the dynamical coordinates. Moreover, they play no further role in the description of the system, nor in the response. Therefore, they will be eliminated at the  stage of the analysis which concerns real physical response.

Finally, the positive definite sector of the Hamiltonian matrix can be symplectically diagonalized into harmonic oscillators, with canonical two-by-two blocks of the form $\Omega\mone_2$, with individual frequencies (energies) $\Omega$.

After this preliminary exposition of the aspects of Williamson's theorem relevant for our purposes, we shall now proceed as follows: first we shall depict the general relationship between the gain-immitance matrix function and the Hamiltonian matrix. Next we shall show the integral formulae for the causal and anticausal propagators, and put them together for an integral expression of the fundamental matrix $\exp(t\mathsf{Jh})$. Once this is given, and considering that under canonical quantization the Heisenberg equations of motion are also solved in terms of the fundamental matrix, we particularize for the case with only oscillators to express the quantum fluctuation--dissipation theorem.

Let the classical dynamical canonical coordinates on phase space be denoted by $\xi^\alpha$, organized in vectors $\bxi$. The classical Hamiltonian is $H=\bxi^T\mathsf{h}\bxi/2$, with $\mathsf{h}$ a positive definite symmetric matrix. The coordinates are canonical, such that $\left\{\bxi,\bxi^T\right\}=\mathsf{J}$.

The evolution of the linear system is governed by $H$. Given that we have a quadratic theory, we can 
study the \emph{exact} linear response; that is, evolution driven by an external force $f(t)$ coupled to a linear combination of the canonical coordinates, $X= \mathbf{x}^T\bxi$, where $\mathbf{x}$ is a column vector of numerical coefficients. That is, evolution guided by the Hamiltonian
\begin{equation}
	\label{eq:hamf}
	H_f= H- f(t) X\,.
\end{equation}
The classical (and Heisenberg) equations of motion in this case are
\begin{equation}
	\label{eq:eomf}
	\dot{\bxi}= \mathsf{J h} \bxi- f(t) \mathsf{J}\mathbf{x}\,.
\end{equation}
Denote the (causal) Laplace transforms with a tilde, as in $\tilde{f}(s)=\int_{0}^\infty\mathrm{d}t\, f(t)e^{-st}$. We then have
\begin{equation}
	\label{eq:Zetas}
	\tilde{\bxi}(s)= \left(s\mone -\mathsf{J h}\right)^{-1} \bxi(0) - \tilde{f}(s)  \left(s\mone -\mathsf{J h}\right)^{-1}\mathsf{J}\mathbf{x}\,.
\end{equation}
In mechanical and electrical contexts, a \emph{gain-immitance} response function relates a \emph{velocity} $\dot{X}$ with a force $f_Y$, $\dot{X} = W_{XY} f_Y$. In electrical circuits in particular, the voltage response to a driving current is called an impedance $\msZ$, while the opposite is known as admittance $\msY$. Voltage and current gains are relations between a drive and an observed quantity with the same nature. In Laplace space, with classical homogeneous initial conditions $\bxi(0)=0$,
\begin{align}
	\label{eq:impedance}
	\widetilde{\left(\dot{X}\right)}&= s\tilde{X}(s)= s \mathbf{x}^T\tilde{\bxi}(s)\nonumber\\
	&= -s\mathbf{x}^T\left(s\mone-\mathsf{J h}\right)^{-1}\mathsf{J}\mathbf{x} \,\tilde{f}(s)\,.
\end{align}

Suppose we have a set of variables $X_i$ with corresponding coefficient vectors $\mathbf{x}_i$, then the response of the linear system in Laplace space will be 
\begin{equation}
	\label{eq:generalizedimmitancematrix}
	\msW_{ij}(s) = - s\mathbf{x}^T_i\left(s\mone-\mathsf{J h}\right)^{-1}\mathsf{J}\mathbf{x}_j\,.
\end{equation}

One can readily realize that the resolvent
$\left(s\mone-\mathsf{J h}\right)^{-1}$ is the Laplace transform of
$\exp\left(t \mathsf{J h}\right)$. Defining the response (gain-immitance) matrix function on the linear symplectic space spanned by the $\xi^\alpha$ coordinates as
\begin{equation}
	\label{eq:Wresponse_mat}
	\msW(s)= - s\left(s\mone-\mathsf{J h}\right)^{-1}\mathsf{J}\,,
\end{equation}
we can solve for the resolvent as
\begin{equation}
	\label{eq:resolvent}
	\left(s\mone-\mathsf{J h}\right)^{-1}= \frac{1}{s}\msW(s) \mathsf{J}\,.
\end{equation}
Now, since the resolvent is the Laplace transform of the fundamental matrix $\exp(t\mathsf{Jh})$, we can invert the Laplace transform, but writing the resolvent in the form of eq. (\ref{eq:resolvent}). For definiteness, we make the step function $\Theta(t)$ explicit to write
\begin{equation}
	\label{eq:causalfundamental}
	e^{t\mathsf{Jh}}\Theta(t)= \frac{i}{2\pi}\int_{\mathbb{R}}\frac{\mathrm{d}\omega}{\omega+i0^+} e^{i\omega t}\mathsf{W}\left(-i\omega+0^+\right)\mathsf{J}\,.
\end{equation}
Following the same procedure for \emph{anticausal} forces, we obtain
\begin{equation}
	\label{eq:anticausalfundamental}
	e^{t\mathsf{Jh}}\Theta(-t)= \frac{-i}{2\pi}\int_{\mathbb{R}}\frac{\mathrm{d}\omega}{\omega-i0^+} e^{i\omega t}\mathsf{W}\left(-i\omega-0^+\right)\mathsf{J}\,.
\end{equation}

We desire to put these two together to obtain an integral formula relating the fundamental matrix at all times to the gain--immitance matrix distribution
\begin{equation}
	\label{eq:gainimmitancedist}
	\mathsf{W}_d(\omega):=\lim_{\epsilon\to0^+}\mathsf{W}(-i\omega+\epsilon)
\end{equation}
In order to do so, first observe that $\msW^{\dag}(s)=-\msW(-s^*)$, which follows from 
\begin{align}
	\msW^{\dag}(s)=s^*\mathsf{J}\left(s^*\mone+\mathsf{hJ}\right)^{-1}\,,\nonumber
\end{align}
where we have used that $\mathsf{h}$ is hermitian and $\mathsf{J}$
is antihermitian. Now, since $\mathsf{J}^2=-\mone$ we see that
$\mathsf{J}\left(s^*\mone+\mathsf{hJ}\right)=\left(s^*\mone+\mathsf{Jh}\right)\mathsf{J}$,
whence
$\left(s^*\mone+\mathsf{Jh}\right)^{-1}\mathsf{J}=\mathsf{J}\left(s^*\mone+\mathsf{hJ}\right)^{-1}$.
Therefore,
\begin{align}
	\label{eq:fundamentalgainone}
	e^{\mathsf{J h}t}=&\frac{i}{\pi}\int_{\mathbb{R}}\mathrm{d}\omega\,e^{-i\omega t}\left[\frac{\msW\left(-i\omega+0^+\right)}{\omega+i 0^+}\right.\nonumber\\
	&\left.-\frac{\msW\left(-i\omega-0^+\right)}{\omega-i 0^+} \right]\mathsf{J}\nonumber\\
	=& \frac{i}{\pi}\int_{\mathbb{R}}e^{-i\omega t}\left\{\frac{\msW_d\left(\omega\right)}{\omega+i 0^+}+\left[\frac{\msW_d\left(\omega\right)}{\omega+i 0^+} \right]^\dag\right\}\mathsf{J}\
\end{align}
This result, eq. (\ref{eq:fundamentalgainone}), is general. Let us now restrict ourselves to the case in which the Hamiltonian matrix $\msf{h}$ is positive definite. Then $\msW(0)=0$, and from here
\begin{align}
	e^{\mathsf{J h}t}	&=\frac{i}{\pi}\int_{\mathbb{R}}\mathrm{d}\omega\,e^{-i\omega t}\frac{1}{\omega+i 0^+} \msW^H(\omega)\mathsf{J}\nonumber\\
	&= \frac{i}{\pi}\int_{\mathbb{R}}\mathrm{d}\omega\,e^{-i\omega t}\mcl{P}\frac{1}{\omega} \msW^H(\omega)\mathsf{J}\,,
\end{align}
where we have introduced the notation $\mathsf{W}^H$ for the hermitian part of $\mathsf{W}_d$,
\begin{align}
	\label{eq:whdef}
	\mathsf{W}^H(\omega)& = \frac{1}{2}\left[\mathsf{W}_d(\omega)+\mathsf{W}_d^\dag(\omega)\right]\\
	&=\frac{1}{2}\left[\msW\left(-i\omega+0^+\right)-\msW\left(-i\omega-0^+\right)\right]\,.\nonumber
\end{align}
Furthermore, the behaviour of $\msW(s)$ close to the origin is linear, and thus the principal part sign can be discarded to yield the final result for positive definite Hamiltonian matrices,
\begin{equation}
	\label{eq:finalfundamental}
	e^{t \mathsf{Jh}}= \frac{i}{\pi}\int_{\mathbb{R}}\frac{\mathrm{d}\omega}{\omega}\mathsf{W}^H(\omega)\mathsf{J}\,.
\end{equation}

Our objective is the quantum fluctuation--dissipation formula. Thus we shall now make use of these results in the quantum context.
Canonical quantization of oscillatory degrees of freedom is achieved by elevating the dynamical phase space coordinates to operators, i.e., the minimal set of descriptive variables, with commutation relations $\left[\bxi,\bxi^T\right]=i \hbar\mathsf{J}$. Generically for linear systems, the classical \emph{and} quantum Heisenberg equations of motion are
\begin{equation}
	\label{eq:eoms}
	\dot{\bxi}= \mathsf{J}\mathsf{h}\bxi\,.
\end{equation}
with solution also given by the fundamental matrix $\exp(t \mathsf{Jh})$, $\bxi(t)=\exp(t \mathsf{Jh})\bxi(0)$.

Therefore, the set of two-point correlators   $\left\langle\bxi(t) \bxi^T(0) \right\rangle$ is determined from the equal time correlator as
\begin{align}
	\label{eq:timecorr}
	\left\langle\bxi(t) \bxi^T(0) \right\rangle &= e^{t \mathsf{J h}}  \left\langle\bxi(0) \bxi^T(0) \right\rangle
\end{align}
Now, having frozen \emph{before quantization} the nondynamical variables, which in our case are an artifact created by the nonreciprocal constraints being expressed in an extended configuration space, and leaving out of the response all the  free-particle dynamics, i.e., no poles at $s=0$ in $\msW(s)$, we can express the two-point correlators by the  general formula
\begin{align}
	&\left\langle\bxi(t)\bxi^T(0)\right\rangle=\frac{i}{\pi}\int_{\mathbb{R}}\frac{\mathrm{d}\omega}{\omega}e^{-i\omega
		t}\msW^H\left(\omega\right)\msJ \left\langle\bxi(0)\bxi^T(0)\right\rangle,\label{eq:corr_general}
\end{align}
where we have made use of (\ref{eq:finalfundamental}). 

Projecting (\ref{eq:corr_general}) on the upper/lower coordinates of the diagonal, and assuming that the harmonic oscillators are in thermal equilibrium we shall now derive the general formulae (\ref{eq:Phit_Phi0_GF}) and (\ref{eq:Qt_Q0_GF}) in the main text (MT) for reciprocal and non-reciprocal multiport linear systems (with a complete description in terms of dynamical variables). In order to achieve this goal, because of linearity, it is enough to show that for each term of the form of eq. (\ref{eq:twoportstage}) in the MT there is a contribution of the relevant form.

Next assume that $\bxi$ are the canonical coordinates, such that the Hamiltonian matrix has been symplectically diagonalized and is block diagonal, with each block being $\hbar\Omega_k\mone_2$. Then the equal time correlation matrix is also block diagonal, with each block being
\begin{equation}
	\label{eq:thermal_block}
	\langle\bxi(0)\bxi^T(0)\rangle_k=\hbar\left[n_{\mathrm{th}}(\Omega_k)+\frac{1}{2}\right]\mone_2-\frac{\hbar}{2}\sigma_y\,.
\end{equation}
It is also the case that $\mathsf{W}^H(\omega)$ is block diagonal, and each block is
\begin{equation}
	\label{eq:whblock}
	\mathsf{W}^H(\omega)_k=\frac{\omega\pi}{2}\left(\delta^-_k\mone_2-\delta^+_k\sigma_y\right)\,,
\end{equation}
where we have defined $\delta^\pm_k=\delta(\omega-\Omega_k)\pm\delta(\omega+\Omega_k)$. We now consider $\mathsf{W}^H(\omega)\mathsf{J} \langle\bxi(0)\bxi^T(0)\rangle$, block diagonal, with blocks
\begin{align}
	\label{eq:int_block}
	I_k&=\left[\mathsf{W}^H(\omega)\mathsf{J}\langle\bxi(0)\bxi(0)^T\rangle\right]_k\nonumber\\
	&=  \frac{\omega\pi}{2}\left(\delta^-_k\mone_2-\delta^+_k\sigma_y\right)\left[i\sigma_y\right]\times\dots\nonumber\\
	&\quad\,\times\left\{\hbar\left[n_{\mathrm{th}}(\Omega_k)+\frac{1}{2}\right]\mone_2-\frac{\hbar}{2}\sigma_y\right\}\nonumber\\
	&= \frac{i\hbar\omega\pi}{4}\left(\delta_k^-\sigma_y-\delta_k^+\mone_2\right)\left[\coth\left(\frac{\beta\hbar\Omega_k}{2}\right)\mone_2-\sigma_y\right]\nonumber\\
	&= \frac{i\hbar\omega\pi}{4}\left[\coth\left(\frac{\beta\hbar\omega}{2}\right)\left(\delta_k^+\sigma_y-\delta_k^-\mone_2\right)\right.\nonumber\\
	&\left.\quad\,+\left(\delta_k^+\sigma_y-\delta_k^-\mone_2\right)\right]\nonumber\\
	&= \frac{i\hbar\omega\pi}{2}\left[n_{\mathrm{th}}(\omega)+1\right]\left(\delta_k^+\sigma_y-\delta_k^-\right)\nonumber\\
	&=- {i\hbar}\left[n_{\mathrm{th}}(\omega)+1\right] \mathsf{W}^H(\omega)_k\,.
\end{align}
Therefore, for canonical coordinates we can express eq. \eqref{eq:corr_general} as
\begin{align}
	\label{eq:canonical}
	\left\langle\bxi(t)\bxi^T(0)\right\rangle_{\mathrm{can.}}&= \frac{\hbar}{\pi}\int_{\mathbb{R}}\frac{\mathrm{d}\omega}{\omega}e^{-i\omega t}\left[n_{\mathrm{th}}(\omega)+1\right]\mathsf{W}^H(\omega).
\end{align}

In conclusion, for  positive definite Hamiltonian matrices and thermal equilibrium for each constituent harmonic oscillator we have obtained the results of eqs. (\ref{eq:Phit_Phi0_GF}) and (\ref{eq:Qt_Q0_GF}) in the canonical case. As the general case is related to the canonical case by a change of coordinates, we can conclude that for linear lossless passive systems in open/short boundary conditions the time evolution of correlations of inner ($\bxi$) variables is also expressed as Eq. (\ref{eq:canonical}), i.e., 
\begin{align}
	\label{eq:general}
	\left\langle\bxi(t)\bxi^T(0)\right\rangle&= \frac{\hbar}{\pi}\int_{\mathbb{R}}\frac{\mathrm{d}\omega}{\omega}e^{-i\omega t}\left[n_{\mathrm{th}}(\omega)+1\right]\mathsf{W}^H(\omega)\,,
\end{align}
where now the gain--immitance matrix is expressed in the variables being used.

To complete the proof we need to relate the ports variables to the inner ($\bxi$) ones. This is achieved by linear transformations, and the impedance and admittance matrices are determined by those same linear transformations, thus completing the proof of eqs. (\ref{eq:Phit_Phi0_GF}) and (\ref{eq:Qt_Q0_GF}) for linear passive lossless multiport sytems. 

Even though this proof  is now complete and general, the concrete computation for the two most relevant examples might prove illustrative for the reader, and we now present those two. First the standard $LC$ oscillator, and next the two-port nonreciprocal example of Fig. 2(a) in MT.
\subsection{Quantum LC oscillator}
The quantum LC circuit is described by the standard Hamiltonian
\begin{equation}
	\label{eq:LCHam}
	H=\frac{\tilde{Q}^2}{2C}+\frac{\tilde{\Phi}^2}{2L},
\end{equation}
whose causal impedance function is $Z(s)=\frac{s/C}{\left(s^2+\Omega^2\right)}$, with frequency
$\Omega=\left(LC\right)^{-1/2}$. The generalized gain-immitance matrix (\ref{eq:Wresponse_mat}) for the canonical form of the Hamiltonian $H_c=\frac{\Omega}{2}\bxi^T\bxi$ is
\begin{equation}
	\label{eq:ZLC}
	\msW(s)=\frac{s}{s^2+\Omega^2}
	\begin{pmatrix}
		\Omega&-s\\ s&\Omega
	\end{pmatrix}\,.
\end{equation}
This canonical form of the Hamiltonian is the symplectic diagonalization of the initial one.  For the initial Hamiltonian of eq. (\ref{eq:LCHam})  it follows  from the
symplectic transformation
\begin{equation}
	\label{eq:symplectic}
	\tilde{\bX}=\begin{pmatrix}
		\tilde{\Phi}\\ \tilde{Q}
	\end{pmatrix}=\mathsf{S} \bxi=
	\begin{pmatrix}
		R^{1/2}&0\\0& R^{-1/2}
	\end{pmatrix}\bxi, 
\end{equation}
where $R=\sqrt{L/C}$. We  compute the flux and charge time correlators with eq. (\ref{eq:corr_general}) above as
\begin{align}
	\label{eq:phiphit}
	\left\langle\tilde{\bX}(t)\tilde{\bX}(0)\right\rangle =&\, \mathsf{S} \left\langle\bxi(t)\bxi^T(0)\right\rangle \mathsf{S}^T
	\nonumber\\
	=&\,
	\frac{i}{\pi}\int_{\mathbb{R}}\frac{\mathrm{d}\omega}{\omega}
	e^{-i\omega t}\mathsf{S}\msW^H(-i\omega+0^+)\mathsf{S}^T \mathsf{J}\nonumber\\ 
	&\times \mathsf{S}\left\langle\bxi(0)\bxi^T(0)\right\rangle \mathsf{S}^T.
\end{align}
The causal part of the generalized response matrix is
\begin{align}
	\msW(-i\omega+0^+)=&\underbrace{\frac{\omega\pi}{2}\left(\mone_2\delta_--\sigma_y\delta_+\right)}_{\tilde{\msW}^H}\nonumber\\ &+\underbrace{i\frac{\omega}{2}\left[\mcl{P}_-\mone_2-\mcl{P}_+\sigma_y\right]}_{\tilde{\msW}^A}, 
\end{align}
where we have defined the  distributions $\delta_\pm=\delta_{\Omega}\pm\delta_{-\Omega}$ ($\delta_{\Omega}=\delta(\omega-\Omega)$), and the combinations of Cauchy principal values $\mcl{P}_\pm=\mcl{P}\frac{1}{\omega-\Omega}\pm\mcl{P}\frac{1}{\omega+\Omega}$. The hermitian part of the internal impedance matrix is given  by just the first two terms. The internal fluctuations at time $t=0$ of the canonical coordinates are of the form of eq. \eqref{eq:thermal_block}. When sandwiching $\bar{\mathsf{W}}^H$ and $\langle\bxi_0\bxi_o^T\rangle$ with $\mathsf{S}$ and $\mathsf{S}^T$ the matrix
\begin{equation}
	\mathsf{SS}^T=\mathcal{R}=
	\begin{pmatrix}
		R&0\\0&R^{-1}
	\end{pmatrix}
	\label{eq:sstisr}
\end{equation}
appears.

Putting everything together, the relevant part of  the integrand of (\ref{eq:corr_general}) reads
\begin{align}
	&i\left(\msS 	\msW^H \msS^T\right)\msJ\left(\msS 	\langle\bxi_0\bxi_0^T\rangle	\msS^T\right)=\nonumber\\
	&\qquad=\frac{\hbar\Omega \pi}{4}\left[\mcl{R}\left(f_\Omega \delta_-+\delta_+\right)-\sigma_y\left(f_\Omega\delta_++\delta_-\right)\right],\nonumber\\
	&\qquad=\frac{\hbar\omega \pi\left(n_{\mathrm{th}}(\omega)+1\right)}{2}\left[\mcl{R}\delta_--\sigma_y\delta_+\right],\nonumber
\end{align}
with $f_\Omega=\coth\left(\tfrac{\hbar\beta\Omega}{2}\right)$, where we recall that $n_{\mathrm{th}}(\omega)=\left(\coth\left(\beta\hbar\omega/2\right)-1\right)/2$ . We recover the result for  flux and charge fluctuations by inserting the value of  its corresponding top and bottom diagonal element in the formula (\ref{eq:phiphit}), obtaining the same as using Eqs. (\ref{eq:Phit_Phi0_GF}) or (\ref{eq:Qt_Q0_GF}), respectively. Cross correlations between the conjugated variables can be computed with the off-diagonal terms, proportional to the gain responses \cite{Nazarov:2009}.

\subsection{Two-port NR quantum harmonic oscillator}
The general formulae for the two-port nonreciprocal harmonic oscillator are trivially computed using the canonical transformation between the internal (quantized) degrees of freedom and the nondynamical coordinates. We present here the case for the circuit in Fig. 2(a) of the MT of a gyrator with capacitors in parallel. The analysis for its dual circuit, in Fig. 2(b), can be trivially performed using as configuration space coordinates the loop charges instead of the node-fluxes. 

The dynamical coordinates   $\tilde{\bX}_1^T=(\tilde{\Phi}_1, \tilde{\Pi}_1)$ and the (virtual) nondynamical $\tilde{\bX}_2$ degrees of freedom, that will be set to zero before quantization, are assembled in a vector of internal variables $\tilde{\bX}$. This is related to the set of  the external variables $\bX$  by
\begin{align}
	\bX=\msf{U}^{-1}\tilde{\bX}=\msf{U}^{-1}\msf{S}\bxi,
\end{align}
where the symplectic transformation specific to the circuit in Fig. 2(a)) of the MT is 
\begin{align}
	\msf{U}^{-1}=\begin{pmatrix}
		1 & 0 & 0 & -R \\
		0 & -R & 1 & 0 \\
		0 & \frac{1}{2} & \frac{1}{2 R} & 0 \\
		\frac{1}{2 R} & 0 & 0 & \frac{1}{2} \\
	\end{pmatrix}.\nonumber
\end{align}
We have also introduced the canonical coordinates $\bxi$, two of which correspond to the dynamical and two to the nondynamical sector. We denote the dynamical canonical coordinates as $\bxi_{\mathrm{dy}}$. 
Note as well the permutation of the coordinates in the definition of the internal and external vectors, $\bX=(\Phi_1,\Phi_2,\Pi_1,\Pi_2)^T$, and $\tilde{\bX}=(\tilde{\Phi}_1,\tilde{\Pi}_1,\tilde{\Phi}_2,\tilde{\Pi}_2)^T$, for convenience in extracting impedance and admittance. The external variable correlators are finally computed by setting  the nondynamical variables to zero,
\begin{align}
	\langle \bX_t \bX_0^T\rangle =\msf{U}^{-1}\msS\begin{pmatrix}
		\left\langle\bxi_{\mathrm{dy}}(t) \bxi_{\mathrm{dy}}^T(0) \right\rangle&0\\0&0
	\end{pmatrix}\msS^T(\msf{U}^{-1})^T\nonumber
\end{align}
Extracting the upper diagonal block matrix with a rectangular matrix $\msf{P}$, we finally obtain the result 
\begin{align}
	\langle\bPhi_t \bPhi_0^T\rangle&=\msf{P}\langle \bX_t \bX^T_0\rangle \msf{P}^T\nonumber\\
	&=\frac{\hbar R}{2}\int_{\mathbb{R}}d\omega \left[n_{\mathrm{th}}(\omega)+1\right] (\mone_2\delta_-+\sigma_y\delta_+)\,,\nonumber
\end{align}
where the hermitian part of the causal impedance matrix can be read from the integrand, matching the one presented in Eq. (\ref{eq:Z_H_2P_NRHO}).

\addcontentsline{toc}{section}{References}

\bibliographystyle{apsrev4-2}
\bibliography{bibliography}

\end{document}